\documentclass[amsmath,amssymb,showpacs,twocolumn]{revtex4}
\usepackage{color}
\usepackage{bm}
\usepackage{graphicx}
\usepackage{epsfig}
\usepackage{amsmath}
\usepackage{amsfonts}
\usepackage{amssymb}
\usepackage{bezier}
\usepackage{color}

\font\tenmib=cmmib10
\font\eightmib=cmmib10 scaled 800
\font\sixmib=cmmib10 scaled 667
\newfam\mibfam

\def\mib{\fam\mibfam\tenmib}
\textfont\mibfam=\tenmib
\scriptfont\mibfam=\eightmib
\scriptscriptfont\mibfam=\sixmib
\mathchardef\alpha="710B
\mathchardef\beta="710C
\mathchardef\gamma="710D
\mathchardef\delta="710E
\mathchardef\epsilon="710F
\mathchardef\zeta="7110
\mathchardef\eta="7111
\mathchardef\theta="7112
\mathchardef\kappa="7114
\mathchardef\lambda="7115
\mathchardef\mu="7116
\mathchardef\nu="7117
\mathchardef\xi="7118
\mathchardef\pi="7119
\mathchardef\rho="711A
\mathchardef\sigma="711B
\mathchardef\tau="711C
\mathchardef\phi="711E
\mathchardef\chi="711F
\mathchardef\psi="7120
\mathchardef\omega="7121
\mathchardef\varepsilon="7122
\mathchardef\vartheta="7123
\mathchardef\varrho="7125
\mathchardef\varphi="7127
\mathchardef\sGamma="7100
\mathchardef\sDelta="7101
\mathchardef\sTheta="7102
\mathchardef\sLambda="7103
\mathchardef\sXi="7104
\mathchardef\sPi="7105
\mathchardef\sSigma="7106
\mathchardef\sUpsilon="7107
\mathchardef\sPhi="7108
\mathchardef\sPsi="7109
\mathchardef\sOmega="710A

\def\nd{^{\vphantom{\dagger}}}
\def\ns{^{\vphantom{*}}}
\def\yd{^\dagger}
\def\frac#1#2{{\textstyle{#1 \over #2}}}

\def\ie{{\it i.e.\/}}

\def\pz{{\partial}}
\def\half{\frac{1}{2}}
\def\bnabla{\boldsymbol{\nabla}}
\def\ij{{\langle ij\rangle}}

\def\beq{\begin{equation}}
\def\eeq{\end{equation}}
\def \be{\begin{equation}}
\def \ee{\end{equation}}
\def \bea{\begin{eqnarray}}
\def \eea{\end{eqnarray}}
\def\half{\mbox{$1\over2$}}

\def\cC{{\cal C}}
\def\cH{{\cal H}}
\def\cD{{\cal D}}

\def\cL{{\cal L}}
\def\cA{{\cal A}}
\def\cP{{\cal P}}

\def\bE{{\mib E}}
\def\bR{{\mib R}}
\def\bP{{\mib P}}
\def\bA{{\mib A}}
\def\bL{{\mib L}}
\def\bV{{\mib V}}

\def\bk{{\mib k}}
\def\bv{{\mib v}}
\def\bj{{\mib j}}
\def\bx{{\mib x}}
\def\br{{\mib r}}
\def\by{{\mib y}}
\def\bz{{\mib z}}
\def\bn{{\mib n}}
\def\bfeta{{\mib\eta}}

\def\nper{n\ns_\perp}
\def\ndot{{\dot n}}
\def\ntil{{\tilde n}}
\def\nhat{{\hat\bn}}

\def\zhat{{\hat\bz}}

\def\bTheta{{\mib\Theta}}
\def\hbOmega{{\widehat{\mib\Omega}}}
\def\bPi{{\mib\sPi}}

\def\sxy{\sigma_{xy}}
\def\rs{{\rm s}}
\def\tV{{\textsf V}}
\def\bRO{{\bR\ns_0}}
\def\bRV{{\bR\ns_{\textsf V}}}
\def\XV{{X\ns_{\textsf V}}}
\def\YV{{Y\ns_{\textsf V}}}
\def\PVa{{P^\alpha_\tV}}
\def\PXV{{P^x_{\textsf V}}}
\def\PYV{{P^y_{\textsf V}}}
\def\PiXV{{\sPi^x_{\textsf V}}}
\def\PiYV{{\sPi^y_{\textsf V}}}
\def\PiZV{{\sPi^z_{\textsf V}}}

\begin{document}

\title{Vortex Dynamics and Hall Conductivity of Hard Core Bosons}
\author{Netanel Lindner$^{1,2}$, Assa Auerbach$^{1}$ and Daniel P. Arovas$^{3}$}
\affiliation{1)  Physics Department, Technion, 32000 Haifa,
Israel}
\affiliation{2) Institute of Quantum Information, California Institute of Technology, Pasadena, CA 91125, USA}
\affiliation{3) Department of Physics, University of California at San Diego, La
Jolla, CA 92093, USA}
\begin{abstract}
Magneto-transport of hard core bosons (HCB) is studied  using an $XXZ$ quantum spin model representation, appropriately gauged on the torus to allow for
an external magnetic field.  We find strong lattice effects near half filling.   An
effective quantum mechanical description of the vortex degrees of freedom is derived. Using semiclassical and numerical analysis we 
compute the vortex hopping energy $t_\tV$, which at half filling is  close to magnitude of the boson hopping energy.   
The critical quantum melting density of  the vortex lattice is estimated at $6.5\times10^{-3}$ vortices per unit cell.
The Hall conductance is computed from  the Chern numbers of the low energy eigenstates.
At zero temperature, it reverses sign abruptly  at half filling.  At precisely half filling, all eigenstates
are doubly degenerate for any odd number  of flux quanta.
We prove the exact degeneracies on the torus by constructing an SU(2) algebra of point-group symmetries, associated with the center of vorticity.  
This result is interpreted as if each vortex carries an internal
spin-half degree of freedom ('vspin'), which can manifest itself as a charge density modulation in its core.  Our findings suggest interesting
experimental implications for vortex motion of cold atoms in optical lattices, and magnet-transport of short coherence length superconductors.
\end{abstract}
\pacs{05.30.Jp, 03.75.Lm, 66.35.+a, 67.85.d}
\today
\maketitle

\section{Introduction}
\label{sec: intro vortex}
Hard core bosons (HCB)  are often used to describe  superfluids
and superconductors which are characterized by low  superfluid stiffness and
short coherence lengths. As such, HCB  are relevant to cold atomic gases
in optical lattices \cite{PS, Zoller}, low capacitance Josephson
junction arrays~\cite{Ood,Altman-JJ,JJReview},  disordered
superconducting films~\cite{fisher}, and cuprate
superconductors~\cite{Uemura,EK-Uemura,Ranninger,LB-cuprates,PBFM,LB2,MA}.

At low densities, HCB can be treated by weak coupling (Bogoliubov) perturbation theory~\cite{FW}.
Closer to half filling, lattice umklapp scattering and the hard core constraints become important.
Recent calculations of the dynamical conductivity of HCB near half filling~\cite{LA} demonstrate the breakdown of
weak-scattering Drude-Boltzmann transport theory in this regime. HCB exhibit so-called 
  `bad metal' phenomenology, (i.e. large resistivity, linearly increasing  in temperature). Such behavior has been often observed in
unconventional superconductors~\cite{BM}.

This paper also concerns dynamical correlations of HCB and their vortices near half filling. These will be exposed by including a weak orbital magnetic field in the Hamiltonian
and studying the Hall effect.

Our primary results are as follows. 
Firstly, we apply a combination of semiclassical analysis and exact diagonalization to the gauged $XXZ$ Hamiltonian on a finite
latice on the torus. We highlight the (sometimes overlooked) fact that
a uniform magnetic field of one flux quantum penetrating the surface of
the torus beaks translational
symmetry. As a consequence, the semiclassical vortex center  is subjected to a confining potential minimized at a well defined position. 
Fitting the low many-body spectrum to an effective single-vortex Hamiltonian, we determine the vortex hopping rate (effective mass).

Near half filling, the
vortex mass is found to be similar in magnitude to the HCB mass.
This allows us to estimate  the critical field for quantum melting of the vortex solid (superfluid) phase at
6.5$\times 10^{-4}$ flux quanta per unit cell.
Secondly, at half filling we find doublet degeneracies associated with an odd number of magnetic flux quanta 
 penetrating the torus.  We associate them with symmetries about the vortex position, and label the emergent degrees of freedom as `vortex spin'
 (v-spin). Physically, these degrees of freedom correspond to the orientation of the charge density wave  in the vortex cores.

Finally, we compute the Hall conductivity using thermally averaged Chern numbers.   In
stark contrast to continuum bosons, and to electrons in metallic bands, we find that the Hall conductivity of HCB reverses sign abruptly at half filling.
The associated Hall temperature scale vanishes at half filling, signaling a  possible quantum phase transition for the thermodynamic system in a magnetic field.  Some of these results were briefly reported in a recent
Letter~\cite{LAA}.

This paper is organized as follows.  In Section \ref{sec:HCB} the HCB Hamiltonian is introduced, with a discussion
of its charge conjugation symmetry about half filling.  Semiclassical approximations are derived in Section \ref{sec:SC}, for the various  regimes of filling.
At low density, we recover the Gross-Pitaevskii theory with its Galilean invariant vortex dynamics
and classical Hall effect.   At half filling, the continuum limit corresponds to  the anisotropic gauged non linear sigma model.
Its vortices possess localized charge density waves in their cores.  Section \ref{sec:GT} describes the mathematical peculiarities of  the gauged torus, including translational symmetry breaking (elaborated in Appendix \ref{App:TSB}). Definitions of  null lines, null points  and vorticity centers are
provided.  The point group symmetry generators $\sPi^x_\tV$ and $\sPi^y_\tV$
are constructed and their commutator is calculated. The  proof of v-spin degeneracies at half filling is provided.
Section \ref{sec:VH} computes the vortex effective hamiltonian by combining  semiclassical and exact diagonalization calculations.
The  critical field for quantum melting of the vortex lattice is deduced from our value of vortex hopping rate.
Section \ref{sec:HALL} computes the Hall conductance on the torus  as a function  of density and temperature. We conclude in
section \ref{sec:EXP} and discuss experimental implications of our results in  cold atoms and  cuprate superconductors.

\label{sec:GT}

\section{Hard Core Bosons}
\label{sec:HCB}
The conventional Bose Hubbard model for interacting bosons is
\bea
\cH\ns_U &=& -2J\sum_\ij\big(e^{iq A\ns_{ij}}\,a\yd_i a\nd_j + a^{-iq A\ns_{ij}} a\yd_j a\nd_i\big)\nonumber\\
&& + 4V \sum_\ij \big(n\ns_i-\half\big)\big(n\ns_j-\half\big) - \mu\sum_i n\ns_i\nonumber\\
&&+\half U\sum_i n\ns_i\big(n\ns_i-1\big)\ ,
\label{BHM}
\eea
In the hard core limit $U\to\infty$, Eq.~(\ref{BHM}) reduces to the HCB Hamiltonian as $\cH=\cP\, \cH\ns_{U=0}\, \cP$,
where $\cP$ is the projector onto the subspace where $n\ns_i=0$ or $1$ for each site.

We use units where $\hbar=c=1$.   $\ij$ denotes a nearest neighbor
link on the square lattice; the lattice constant is $a=1$. $J$ is
the Josephson coupling, $q$ is the boson charge and $A\ns_{ij}$
the electromagnetic gauge field on a bond.  $V$ is a nearest
neighbor repulsive interaction. In the HBC limit,  The chemical
potential $\mu=0$ corresponds to a density of half filling
$\langle n \rangle=\half$, with half a boson per site on average.

As is well-known, HCB operators obey an algebra corresponding to spin-$\half$:
\bea
{\tilde a}\yd_i &=& \cP \, a\yd_i \, \cP = S^+_i \nonumber\\
{\tilde a}\nd_i &=& \cP \, a\nd_i \, \cP = S^-_i \nonumber\\
{ n}\nd_i &=& {\tilde a}\yd_i {\tilde a}\nd_i = S^z_i+\half.\label{HCB-CR}
\eea
By $[S^+_i,S^-_j]=2S^z_i\,\delta\ns_{ij}$, HCB operators obey constrained commutation relations,
\bea
\big[{\tilde a}\nd_i  \, , \, {\tilde a}\yd_j\big]=\big(1-2{  n}\ns_i\big)\,\delta\ns_{ij}\ .
\eea
The constraint effects of  $-2{ n}\ns_i\delta_{ij}$ become important near half filling.
$\lim_{U\to \infty} \cH_U$ is thus represented by the gauged spin-half quantum $XXZ$ model,
\bea
\cH&=&  -2J  \sum_\ij
\left(  e^{iqA\ns_{ij} } S^+_i S^-_j + e^{-iqA\ns_{ij}} S^-_i S^+_j \right)\nonumber\\
&&\qquad +4V \sum_{\langle i,j\rangle}    S^z_i S^z_j - \mu  \sum_i (S^z_i+\half)  .
\label{xxz}
\eea
It is widely believed that the ground state of Eq.~(\ref{xxz})
exhibits magnetic order.  In the
regime of  $V\ll J$, which is relevant to this paper,  the ordered moment lies in the $XY$
plane, $\langle S_i^+\rangle \ne 0$. That is to say,  except for the limits $n=
0,1$,  the ground state of HCB exhibits long range superfluid order.

\subsection{HCB charge conjugation symmetry}
Another important distinction between the HCB Hamiltonian (\ref{xxz}) and the finite $U$ Bose-Hubbard model of Eq. (\ref{BHM}), is the emergence of
charge conjugation symmetry in the infinite $U$ limit.  One defines the unitary charge conjugation operator,
\be
C  \equiv  \exp\Big(i\pi  \sum_i S^x_i \Big).
\label{Part-hole1}
\ee
$C$ transforms``particles'' into ``holes'', \ie\ $C\yd{\tilde n}\ns_i C = 1-{\tilde n}\ns_i$, and
\be
C^\dagger\, \cH\big(q\bA, \mu\big)\,C  =   \cH\big(-q\bA, -\mu\big)\ .
\label{Part-hole}
\ee
At half filling ($\mu=0$), and $\bA=0$,  the Hamiltonian is invariant under  charge conjugation {\em on any lattice structure}~\cite{comm:CCS}.

A consequence of (\ref{Part-hole}) is that the Hall conductivity  (which is linear in $q$) is {\em antisymmetric}
in the deviation from half filling, ie
\be
\sigma_{xy}(n,T)  = - \sigma_{xy}(1-n,T).
\label{phs}
\ee
In contrast,  the superfluid stiffness $\rho_s(n)$ and longitudinal conductivity $\sigma_{xx}(n)$ are symmetric
under $n\to (1-n)$.

In terms of vortex motion, (\ref{phs}) implies that below and above half filling vortices drift in opposite
directions relative to the particle current.

\section{Semiclassical theory}
\label{sec:SC}
The partition function of HCB can be represented by the spin half coherent state path integral~\cite{Book,Notes},
\be
Z= \int \!\cD \hbOmega(\tau) \exp\left (  \int_0^\beta d\tau \left( iK - H^{cl}\right)\right),\label{Z}
\ee
where
\bea
K[\hbOmega,\dot{\hbOmega}]&\equiv &\half  \sum_i (1-\cos\theta_i)\,\dot{\phi}_i \\
H^{\rm cl}[\hbOmega,\bA]&=&  - J \sum_{\langle i,j \rangle}   \sin\theta_i\sin\theta_j \cos\left( \phi_i-\phi_j + q A_{ij}\right) \nonumber\\
&& ~+V \sum_{\langle i,j \rangle} \cos\theta_i \cos\theta_j -  {\mu\over 2} \sum_i \cos\theta_i .
\label{CSPI}
\eea
  $\hbOmega_i=(\theta_i,\phi_i) $ are the polar angles on a sphere. The spin size $S=\half$
plays the role of the large parameter which controls the semiclassical expansion.

In the classical  (saddle point) approximation, for $\bA=0$,  the  ground state superfluid stiffness  is
\bea
\rho_{\rm s}^{\rm cl} &=&   q^{-2} {\pz^2\!  H^{\rm cl}\over \pz A^2_{\br,\br+\hat{\bx}}}\Bigg|_{\bA=0}\nonumber\\
&=&  J\, \big\langle \sin\theta_\br \big\rangle = 4J n(1-n).
\label{rho-cl}
\eea
which (in contrast to continuum bosons)  exhibits a non-monotonic dependence on $n$.  At half filling (optimal density), 
$\rho\ns_{\rm s}$ is maximized. Quantum corrections enhance  $\rho_{\rm s}^{\rm cl}(n=\half)$ further by about 7\%~\cite{sandvik,Troyer}.
The superfluid stiffness vanishes at the Berezinskii-Kosterlitz-Thouless (BKT)~\cite{BKT} transition temperature, computed to be  $T\ns_{\rm BKT}\simeq 1.41 J  $.

The kinetic term $K$ of Eq.~(\ref{Z}), determines the quantum dynamics.
The harmonic spin-wave expansion of  (\ref{CSPI}) yields a linearly dispersing phase fluctuations mode. The  order parameter is suppressed to zero at all finite temperatures,
 in accordance with the Mermin-Wagner theorem.

\subsection{Low density, Gross-Pitaevskii limit}
For large negative values of the chemical potential $\mu$, the action in Eq.~(\ref{CSPI}) can be expanded
around the ferromagnetic (low density) state  of $\theta_i\approx \pi$,  
\be
\cos\theta_i \to 2n_i-1 \quad,\quad \sin\theta_i \approx 2\sqrt{n_i(n_i-1)}\ .
\ee
We define the continuous field
\be
\psi(\bx_i)= \sqrt{n_i\over a}\>e^{i\phi_i},
\ee
where $a$ is the lattice constant, and replace  the measure by
\be
\prod_i\cD \cos\theta_i \,\cD\phi_i \longrightarrow   \cD\psi^*\,\cD\psi \prod_{i,t} \Theta\! \left(1 -\int\limits_{V_i}\! \!  d^2\! x\> \psi^*\psi \right)  \ ,
\ee
up to an unnecessary normalization constant.  The Heaviside functions enforce the hard core constraint $n\ns_i \le 1$ in the $i^{\rm th}$ unit cell at  each time slice.
In the low density limit, these constraints are ignored, and the action (\ref{CSPI}) is expanded
to leading order in $n_i\ll \half $, and gradients $\nabla \psi$.  This yields an effective Gross-Pitaevskii (GP) theory~\cite{PS},
\bea
Z\ns_{\rm GP}&= &\int \!\cD\psi^*\,\cD\psi \> \exp \big( -  S\ns_{\rm GP}[\psi^*,\psi,\bA] + \ldots \big)\nonumber\\
S\ns_{\rm GP}&=&  \int \!\!d^2 x \! \int \!\!dt \>\Big[  \psi^* (\partial_t -\mu) \psi   \nonumber\\
&&  \qquad+ {1 \over 2 m^*} \big| (-i\nabla - q \bA)\psi\big|^2 + \half g |\psi|^4 \Big]
\label{GP}
\eea
where the effective mass and interaction parameters are given by
\bea
 m^*  &=& {1\over 16 J}\nonumber\\
 g &=& 16(J+V) .
\eea
 
 In the presence of a magnetic field $B\hat{\bz}$,  a density of $n_v=B/\phi\ns_0$ vortices is produced, where $\phi\ns_0=2\pi/q$ is the
flux quantum.  The core profile function $f(\br-\bR_j)$ near vortex $j$ is well approximated by minimizing the GP energy, which
yields~\cite{PS}
\be
f\ns_{\rm GP}(\br) \simeq  { \sqrt{n} \>r  \over \sqrt{ \xi^2+ r^2 }}\ ,
\label{GP-profile}
\ee
where $\xi = 1/\sqrt{g m^* n}$ is the coherence length.  For $n\ll\half$, one has $\xi\gg a$.
The core density depletion is proportional to $1-|f_{\rm GP}|^2$. Hence it decays as $1/r^2$ away from the vortex center.

In the high density limit, $n\to  1$,  the partition function
can also be approximated by the same GP action (\ref{GP}) following  
a particle hole transformation (\ref{Part-hole}).
In this case, $|\psi|^2$  represents the density of holes.

By neglecting the higher order gradients, and the hard core constraints, the
 GP theory does not include lattice scattering effects as  it is completely Galilean invariant.
Consider an externally induced uniform current density
\be
\bj=qn  \bv_s.
\ee
In the  moving frame
of velocity $\bv_s$ the  vortices are stationary. Therefore, back in the lab frame, a purely transverse electromotive field is produced by the moving vortices,
\bea
\bE &=& {h\over q} \hat{z} \times \bj_{\rm v}\nonumber\\
&=& {h\over q}n_v   \hat{z} \times\bV_{\rm v}\nonumber\\
&=& {h\over q^2}{ n_v\over n}   \zhat \times\bj .
\label{Lorentz}
\eea
That is to say, in the pure GP theory, the longitudinal (dissipative) conductivity vanishes, and  the Hall conductivity
equals to the classical value,
\bea
\sigma_{xx}&=&0\nonumber\\
 \sigma_{xy} &=&  \left({q^2 \over h}\right)  \left( {n\over n_v}\right)= {nq \over B}\ .
 \label{GI-sxy}
 \eea

Spoiling Galilean invariance by the presence of nonuniform potentials, boundary conditions, or by an underlying lattice
can allow vortices to tunnel between different real space positions, resulting in a longitudinal conductivity~\cite{AAG,AA}.

\subsection{Half filling, anisotropic $\sigma$-model}
Toward half filling, lattice scattering  modifies the vortex  structure and dynamics.
At half filling $\mu=0$,  the semiclassical  theory of  Eq.~(\ref{CSPI})  is described by
the anisotropic Non Linear $\sigma$-Model (NLSM) \cite{haldane88}.  After a sublattice rotation $\prod_{i\in B} e^{i\pi S^z_i}$  all the pseudo-spin interactions are antiferromagnetic.  The spins $\hbOmega_i$  are represented by
\be
\hbOmega_i =  \eta\ns_i\, \nhat(\bx\ns_i)\, \sqrt{ 1- \big(\bL(\bx\ns_i)/S\big)^2 } + \bL(\bx\ns_i)/S\ ,
\label{spins}
\ee
where $\eta\ns_i=\pm 1$  on the $A$ $(+)$ and $B$ $(-)$ sublattices, respectively.  The N\'eel vector $\nhat$ satisfies $\nhat^2=1$
and is orthogonal to the local magnetization $\bL$, \ie\ $\nhat\cdot\bL=0$.

The complex combination
\be
\nper=n\ns_x+in\ns_y=|\nper|\,e^{i\phi},
\label{tn}
\ee
defines the local superfluid order parameter, and
$n_z$ corresponds to a bipartite charge density wave (CDW) with two possible signs.
Following Refs. \cite{haldane88,Book,Notes},   we substitute Eq.~(\ref{spins}) in the measure and action of (\ref{CSPI}),
and expand them to quadratic  order  in $\bL$ and $\nabla \nhat$.
Integrating out $\bL$  arrives at the anisotropic  NLSM path integral,
\be
Z\ns_{\rm NLSM}=\int\!\cD\nhat\,e^{i\Upsilon[\nhat]}\,e^{-S\ns_{\rm E}[\nhat,\bA]}\ ,
\ee
where $S\ns_{\rm E}=\int\limits_0^\beta\!\!d\tau\!\int\!d^2\!x\,\cL\ns_{\rm E}$ is the Euclidean action, with
\bea
\cL\ns_{\rm E}&=&\half\chi\ns_\perp\,\big| \ndot\ns_\perp \big|^2  + \half \chi\ns_z\, \ndot_z^2\label{nlsm}\\
&&\quad + \half\rho\ns_\rs \big| (\bnabla - iq\bA)\,\nper \big|^2 + \half \rho^z_\rs \,(\bnabla n\ns_z)^2 + m_z^2\,n_z^2, \nonumber
\eea
and
\be
\Upsilon=  S \sum_i \eta\ns_i \int\limits_0^\beta\!d\tau \ (1-n_z)\, {\dot\phi}\ns_i ,
 \ee
is the contribution from the geometric phase.  The bare coupling constants are obtained directly from $H^{\rm cl}$:
\be
\chi\ns_\perp =   {S^2\over 8 J}  \quad,\quad  \chi_z={S^2 \over 4(J+V)} ,
\label{bare}
\ee
and
\be
\rho_s=J \quad,\quad \rho_s^z=V \quad,\quad m^2_z=   2(J-V)\ .
\ee
For $\bA=0$, the isotropic (Heisenberg) limit  is at $J=V,m_z=0$. The N\'eel ground state implies degeneracy between superfluid and CDW order, and the existence of two massless Goldstone modes.
At finite $XY$ anisotropy, $m_z>0$, and there is one massless (phase) mode,
and a gapped CDW  (roton) mode at the CDW ordering wavevector $(\pi,\pi)$.


\begin{figure}[!t]
\begin{center}
\includegraphics[width=7cm,angle=0]{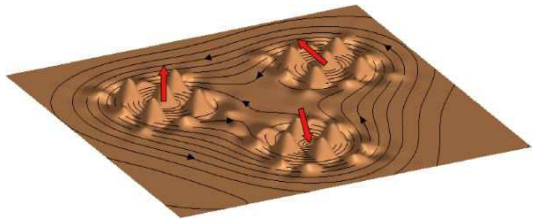}
\caption[Illustration of three static vortices with their v-spins]{Illustration of three static vortices with their v-spins.  Arrows describe the v-spin directions,
their $z$-component is the local charge density wave near the
vortex core, illustrated as ripples in the surface. Current
density is  depicted by black field lines.}
\label{fig:vspin}
\end{center}
\end{figure}


Vortex configurations  at half filling  can be viewed as a localized meron (half skyrmion)  of  the N\'eel field. Since
$|\ntil_{xy}| =0$ at the vortex center,  and $n^2_z=1-|\ntil_{xy}|^2$,
the semiclassical vortex has a CDW in its core, as illustrated in Fig. \ref{fig:vspin}.

Due to the finite  anisotropy `mass'  $m_z>0$, $n_z(\br)$ decays exponentially away from the center
\bea
\nper(\br)  &=&  \sqrt{1-n_z^2(\br)}\ e^{i\phi(\br)} ,\nonumber\\
n_{z}(\br) &\sim& e^{-r/\xi_z} ,\nonumber\\
\xi_z&=&\sqrt{\rho_\rs^z}/m_z .
\label{HF-profile}
\eea
Indeed variational calculations have previously  shown that at half filling  CDW ordering is found
 in the  localized vortex core \cite{SCvortex}.  In   Section \ref{sec:GT}  we shall show that the `orientation' of the charge
density wave is actually a continuous SU(2) symmetry of the quantum Hamiltonian at half filling, which we name {\it v-spin\/}.

Since the system is charge conjugation symmetric at half-filling, there
is no net charge depletion associated with the vortex core, and
thus the statistical Berry phase for exchanging two vortices is
zero.  In other words, the vortices exhibit mutual {\em Bose}
statistics. This is to be contrasted with GP vortices at low filling.
As shown in (\ref{GP-profile}), GP vortices  involve  a  large density
depletion (or accumulation, above half filling), which decays slowly away from their core~\cite{HaldaneWu}.

In the limit where the number of lattice sites $N$ tends to infinity, the confining potential on the vortex vanishes, and
the vortex energy is periodic on the lattice. Its minima lie in plaquette centers (\ie\ at dual lattice sites).

When a vortex moves between dual lattice sites, the path dependent geometric phase $\Upsilon$ yields
$2\pi$ times the number of bosons enclosed by the path.  At half filling, this amounts to an effective $\pi$ flux per dual plaquette.
These phases can be incorporated in an effective hopping model by the dual lattice gauge field $\cA\ns_{\bR,\bR+\bfeta}$
along the link from site $\bR$ to $\bR+\bfeta$.
Thus for a single vortex on the infinite lattice, one can write an effective Harper Hamiltonian,
 \bea
&&H_\tV^\infty  =  -\half t\ns_\tV  \sum_{\bR,\bfeta} \left( e^{i
\cA\ns_{\bR,\bR+\bfeta} } \,  b\yd_{\bR} b\nd_{\bR+\bfeta} + \mbox{H.c.}\right)\nonumber\\
&& \mathop{\sum\hskip-4.9mm \odot\ }_\cC\>\cA\ns_{\bR,\bR+\bfeta}= 2\pi\!\!\! \sum_{i\in {\rm int}(\cC)} \!\! n_i ,
\label{Hv}
\eea
where the sum on the second line is a over a set of links comprising a closed path $\cC$ on the dual lattice, and ${\rm int}(\cC)$ is
the interior of this path, which consists of a set of sites on the original lattice bounded by $\cC$.

\section{The Gauged Torus}
\label{sec:GT}
We now return to the original HCB Hamiltonian,  Eq.(\ref{xxz}).  We consider  a finite 
square lattice, of dimensions $L\ns_x\times L\ns_y$, with $N=L\ns_x L\ns_y$ sites and periodic boundary conditions in
both the $x$ and $y$ directions.  This toroidal geometry is convenient for the study of finite lattices as it minimizes the effects of  boundaries.
It also provides external control over the positions of vortices via the two Aharonov-Bohm (AB) fluxes which run along the two cycles of
the torus.  The lattice site positions are labelled as
\bea
x_i &=& 0,1,2 ,\ldots ,  L_x-1 \nonumber\\
y_i&=& 0,1,2 ,\ldots , L_y -1.
\eea
A uniform magnetic field $B$ is everywhere perpendicular to the surface, such that the total number of flux quanta penetrating the
surface is $N\ns_\phi=NB/\phi\ns_0$, where $\phi\ns_0=2\pi/q$ is the flux quantum.

We construct a (piecewise differentiable)  gauge field $\bA(\bx)$ which interpolates
the lattice gauge field  on the surface of the torus, and obeys
\bea
\hat{\bz}\cdot \nabla \times \bA &=& B \nonumber\\
\bfeta\cdot\bA(\br+\half\bfeta)  &=&   A_{\br,\br+\bfeta}  \ ,
\eea
where $\bfeta=\pm{\hat\bx}$, $\pm{\hat\by}$.
$\bA$ determines the magnetic  fluxes  which flow
through   vertical and horizontal  circumferences of the torus. These are given by the
gauge invariant Wilson loop functions,
\bea
W\ns_y(x) &=& q\oint \!dy\> A\ns_y(x,y)~\mod  2\pi \nonumber\\
W\ns_x(y)&=& q\oint \!dx\> A\ns_x(x,y)~\mod  2\pi .
\label{Wilson}
\eea
The dimensionless AB parameters $\bTheta=(\sTheta\ns_x,\sTheta\ns_y)$
are defined by the Wilson loops at $x=0$ and $y=0$,
\bea
\sTheta\ns_y&=& W\ns_y(x=0) \nonumber\\
\sTheta\ns_x &=& W\ns_x(y=0) .
\eea
$\bTheta$ lives on the reciprocal torus $[0,2\pi )\times[0,2\pi)$.


\begin{figure}[!t]
\begin{center}
\includegraphics[width=8cm,angle=0]{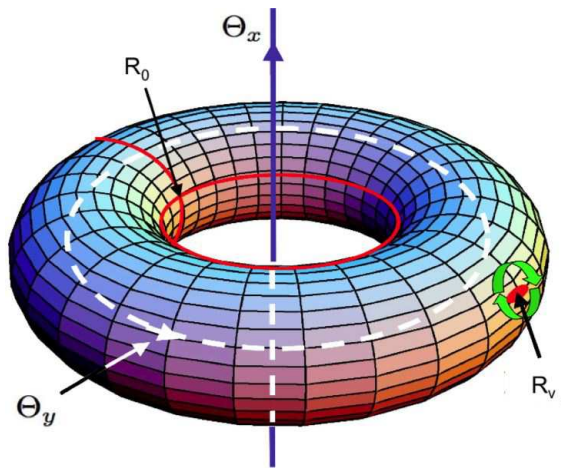}
\caption[The gauged torus]{The gauged torus, which defines the finite size geometry of the HCB.
The torus surface  is penetrated by a uniform magnetic field of
one total flux quantum, and threaded by two Aharonov Bohm fluxes
$\bTheta=(\Theta^x,\Theta^y)$.   Thick (red online)  circles denote the null lines which enclose zero flux, and intersect at the null point $\bR_0(\bTheta)$.
The vorticity center $\bRV$ is located
on the antipodal point to $\bR_0$. The circulating currents of the ground state are illustrated by thick green arrows.
 This geometry  is used  to compute the vortex mass and Hall conductivity of HCB, and to prove the v-spin degeneracies at half filling.}
\label{fig:torus}
\end{center}
\end{figure}

$W\ns_\alpha=0,\alpha=x,y$  define the  {\em null lines} on the torus.
For $N_\phi=1$, there is  one null line in each direction  $x=X_0$, and $y=Y_0$, as depicted   in Fig.~\ref{fig:torus}.  Their intersection is the
{\em null point} $\bRO=(X\ns_0,Y\ns_0)$, which constitutes a gauge invariant symmetry point on the torus.
\bea
X\ns_0(\bTheta) &=& - {L\ns_x \,\sTheta\ns_y\over 2\pi}, \nonumber\\
Y\ns_0(\bTheta) &=&  +{L\ns_y  \, \sTheta\ns_x\over 2\pi} .
\label{eq:vcenter}
\eea
The existence of a special point $R_0$ on the torus, demonstrates
the unintuitive fact that a {\em uniform} magnetic  field
necessarily destroys lattice  translational symmetry.  This fact
is closely related to the quantization of  Dirac  monopoles in
three  dimensions. We elaborate further on this fact in Appendix
\ref{App:TSB}.  Eq.~(\ref{eq:vcenter}) shows that $\bR_0(\bTheta)$
can be moved  continuosly on the torus by changing the AB
parameters $\bTheta$~\cite{HR85,ABHLR88}.

As we shall see in Sec.~\ref{sec:VH},  semiclassical analysis and exact diagonalizations
find that  the center of  vorticity $\bRV$  is located at the {\em antipodal} position of the null point on the torus,
\be
\bRV(\bTheta)= \big( \half L\ns_x + X\ns_0(\bTheta) \, , \, \half L\ns_y + Y\ns_0(\bTheta) \big)\ .
\label{Rv}
\ee

For larger magnetic fields, $N_\phi>1$,  there are $N_\phi$ null lines in each of the $x$ and $y$ directions.
This introduces a set of  $N_\phi^2$ null points which form
an evenly spaced square lattice (which may or may not coincide with the original lattice sites).
These are indexed  by $m,n=0,\ldots, N_\phi-1$
\be
\bR_0^{mn}(\bTheta) = \bR_0(\bTheta) +{1\over N_\phi} (mL_x \,,\, nL_y),
\ee
Correspondingly  there are
$N_\phi^2$ vorticity centers,
\be
\bR_\tV^{mn} (\bTheta)= \bR^{mn}_0(\bTheta)+\half (L\ns_x \,,\, L\ns_y)\ .
\label{Rvmn}
\ee

\subsection{Choosing a gauge}


\begin{figure}[!t]
\begin{center}
\includegraphics[width=10cm,angle=0]{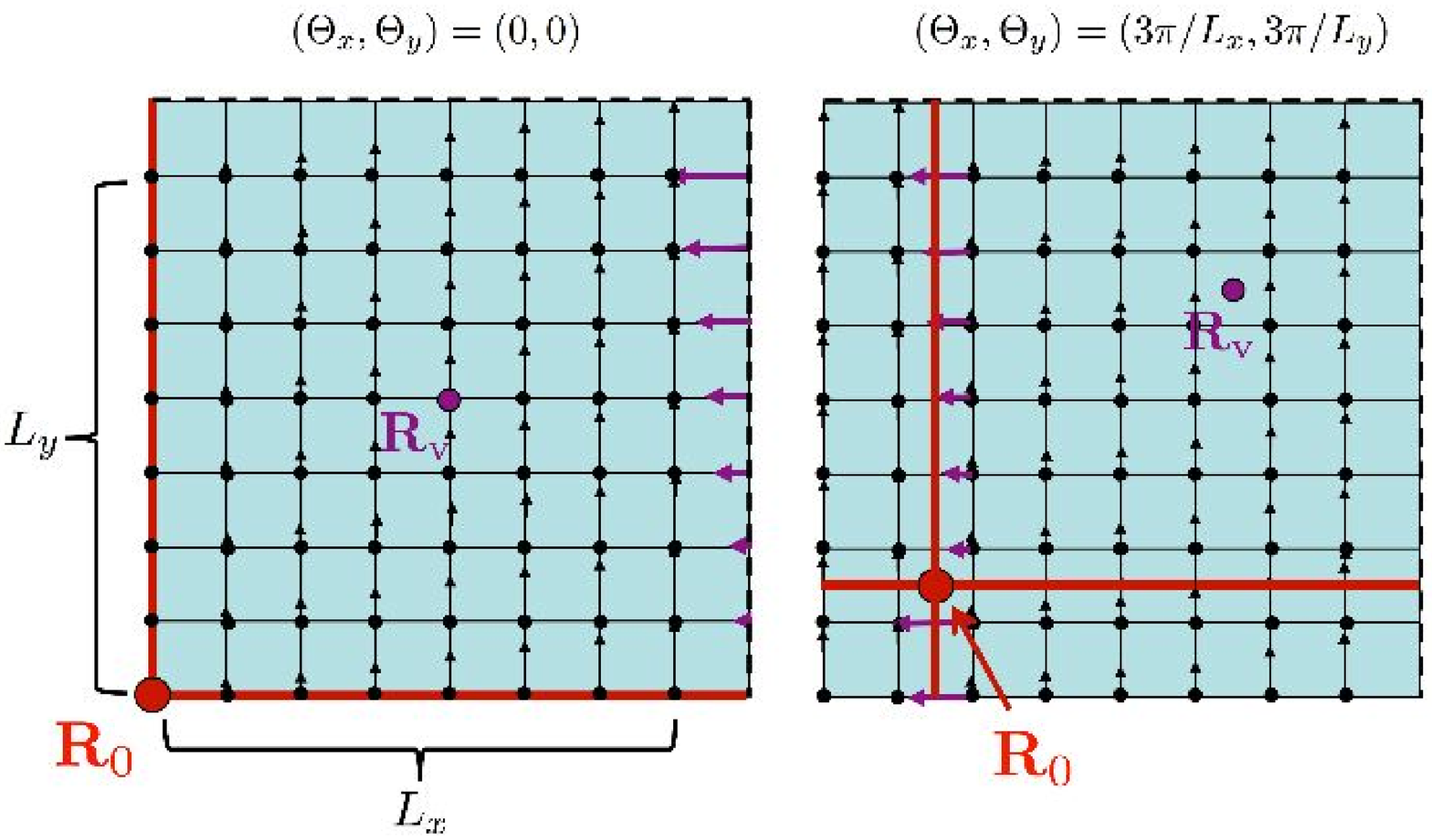}
\caption{The lattice gauge field on links, $\bA_{\br,\br+\bfeta}$, according to Eq.(\ref{eq: gauge}), for two choices of AB parameters
$(\Theta_x,\Theta_y)$.  The length and thickness of the arrows  are proportional to the magnitude of $\bA$.
For a single flux quantum, $N_\phi=1$, the two null lines are marked by red lines.
The null points $\bR_0(\bTheta)$ and vorticity centers $\bR_\tV(\bTheta)$ are depicted.
Note that in the right figure,  both points are located at plaquette centers. }
\label{fig:gauge}
\end{center}
\end{figure}

The uniform magnetic field must integrate to an integer number of flux quanta $N_\phi$,
\be
B={ 2\pi N_\phi \over q L_x L_y}
\ee

The gauge field is given by
 \bea
A^x_{\br,\br+\hat{\bx}} &=&-\mod(y-Y_0,L_y) B \,   L_x \, H(X_0,x) \nonumber\\
 A^y_{\br,\br+\hat{\by}}&=& \mod(x-X_0,L_x) B  \ .
\label{eq: gauge} \eea

Note that for $0<m<n$, $\mod(-m,n)=n-m$. The function $H(X_0,x)$
ensures that $A^x_{\br,\br+\hat{\bx}}$ vanishes unless $\br$ is
immediately to the left ($-\hat{\bx}$) of the null line $x=X_0$.
It is defined by
\be
H(X_0,x) = \left\{ \begin{array}{ll}
1 & \textrm{$\qquad 0<\!\!\mod(X_0-x,L_x)\leq1$}\\
0 & \textrm{$\qquad$ otherwise.}
\end{array} \right.
\ee
For a continuous position $\br$ we define $\bA^\alpha(\br)$ to be
the  linearly interpolated gauge field between the two enclosing
links in the $\alpha$ direction.

 For $\bTheta=(0,0)$ the null point $\bRO$
is at $\bRO=(0,0)$, and the vorticity center is therefore at
$\bRV=(\half L_x, \half L_y)$.

Our  gauge choice is shown in Fig.~\ref{fig:gauge}. The gauge invariant content of $\bA$ consists of the
uniform magnetic field with flux $B=N_\phi\phi\ns_0/L_xL_y$ in each plaquette, and the two Wilson loop functions
 \bea
W\ns_y(x) &=&    x qB L_y +\Theta\ns_y , \nonumber\\
W\ns_x(y) &=&  - y qB L_x +\Theta\ns_x.
\label{eq: wilson loops}
 \eea

\begin{figure}[!t]
\begin{center}
\includegraphics[width=8cm,angle=0]{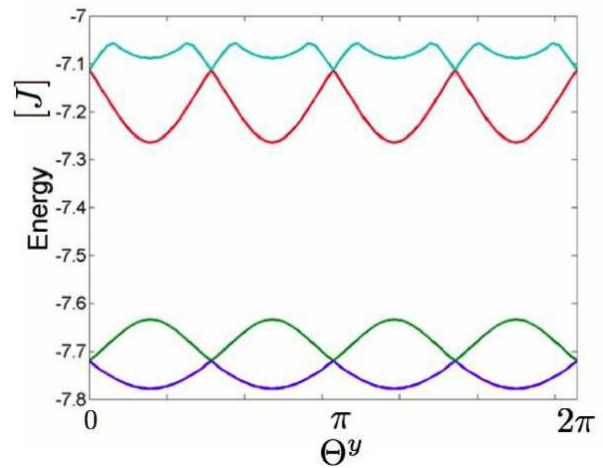}
\caption{Emergence of 'v-spin' degeneracies.
Four lowest eigenenergies of HCB on the torus,  with $N_\phi=1$,   as a function of the AB parameter $\Theta^y$, with
$\Theta^x=0$. The spectrum separates into doublets.
Notice the exact degeneracies which occur when
the vorticity center coincides with a lattice position. } \label{fig:multiplets}
\end{center}
\end{figure}

\subsection{V-spin degeneracies}
\label{sec: Degeneracies}
In the process of  calculating the Hall conductance (see Section \ref{sec:HALL}),
we computed the spectrum at half filling, for an even number of sites, with one total flux quantum of magnetic field.
We encountered  a sequence of  AB fluxes  $\bTheta_i$, where the whole spectrum becomes  two-fold degenerate. These degeneracy points are demonstrated in Fig.~\ref{fig:multiplets} for $N_\phi=1$,
for the lowest two multiplets.
The level crossings indicate the existence of a non-commuting symmetry generators \cite{TH}, which act on
the wavefunctions  of vortices introduced by the external magnetic field.
We  now construct  these  symmetry operators and compute their commutation relations.

As discussed earlier,  for a finite magnetic field ($N_\phi>0$),
$\cH$ does {\em not} possess the lattice translational symmetry.
Nevertheless, with respect to the vorticity center $\bRV(\bTheta)$
we can define  two reflection operators

\bea
\PXV(x,y)&=& (\!\!\!\!\!\!\mod(2\XV-x,L_x),y) ,\nonumber\\
\PYV(x,y) &=& (x,\!\!\!\!\!\!\mod(2\YV-y,L_y)), \eea 
which by
(\ref{Rv}) are equivalent to reflections about $\bR_0$.

Now,  by appropriately tuning $\bTheta$ using
Eq.~(\ref{eq:vcenter}), the vorticity center $\bRV$ can be  chosen
to coincide  with a symmetry point of the square lattice, such as
any lattice site, bond center or plaquette center. Reflecting the
Hamiltonian  about that symmetry point, leads to
 \bea
\PVa \> \cH[\bA] \> \PVa &=&  \cH[\tilde{\bA}] , \nonumber\\
{\tilde \bA}^\alpha_{\br,\br+\bfeta} &=& \bA_{\PVa\br,\PVa(\br+\bfeta)} .
\eea
The gauge invariant content of $\tilde{\bA}^\alpha$  describes  an
inverted uniform magnetic field $\tilde{B}=-B$, and  a reversed
sign of the Wilson loop functions (\ref{Wilson}).

The reversal of the fields in $\tilde{A}$ can be undone, at half filling,  by applying the charge conjugation transformation $C$ (\ref{Part-hole1}), and a pure gauge transformation $U^\alpha$.  Thus, we  construct two  operators,
\bea
\PiXV &=& U^x \, C \, \PXV
\nonumber\\
\PiYV &=& U^y \, C \, \PYV \label{Pialpha}
\eea


\begin{figure}[!t]
\begin{center}
\includegraphics[width=8cm,angle=0]{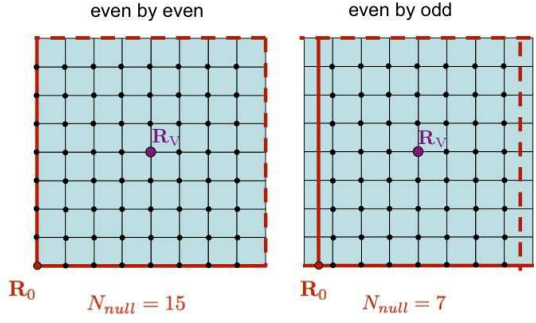}
\caption{ Figuring out $N_{null}$ for two cases of  even size latices.
The parity of the number of lattice sites (black dots) which lie precisely on the null lines  (red color online),
determines the commutation rule of $\PiXV$ and $\PiYV$, according to Eq.~(\ref{Ups-Nnull}). The two examples  explain
why $N_{null}$ is odd whenever  the vorticity center $\bR_\tV$ is positioned precisely on a lattice site.}
\label{fig:Nnull}
\end{center}
\end{figure}


where, \be U^\alpha = \exp\Big(i\sum_\br \chi^\alpha(\br)\,
S^z_\bR\Big)\ , \ee
and
\be
\chi^\alpha(\br)=\int\limits_{\bR_0}^{\br}\!\!d\br'\cdot\left(\bA(\br')+\tilde{\bA}^\alpha(\br')\right) .
\label{chi-alpha}
\ee
In the line   integral we use the interpolated gauge field defined after  Eq.~(\ref{eq: gauge}).
Since $\bA$ and $-\tilde{\bA}^\alpha$ describe the same magnetic fields, they obey,
\be
\nabla \times \left( \bA+\tilde{\bA}^\alpha \right) =0 .
\ee
This implies that $\chi^\alpha$ is independent of which continuous path  (of  zero winding number) is chosen between  $\bR_0$ and $\br$.

It is easy to verify by this construction that for all $\bTheta_i$ such that $\bR\ns_\tV(\bTheta\ns_i)$ is a symmetry point
the $\sPi^\alpha_\tV$ operators become symmetries of the Hamiltonian:
\be
\big[ \cH[\bA] \,,\, \sPi^\alpha_\tV\big] =0 \quad,\quad \alpha=x,y .
\ee

Now we calculate the commutation relation between $\PiXV$ and $\PiYV$. This   is a straightforward but slightly tedious
procedure.   Using the gauge choice (\ref{eq: gauge}), and  (\ref{chi-alpha}),
\bea
\chi^x(\br) &=&  {2\pi N_\phi\over L_y} \mbox{mod}\left(y-Y_0,L_y \right) \left(1 - \delta_{x,X_0}\right), \nonumber\\
\chi^y(\br) &=& 0
, \label{chi} \eea
Note that
$\chi_\tV^x$ vanishes on the null lines. Multiplying the two
$\sPi^\alpha_\tV$ operators yields

\bea
\PiYV\, \PiXV &=&  \exp\Big(i \sum_\br \big( \chi^y- \chi^x(\PYV[\br])\big) S^z_\br\Big)\PYV \, \PXV ,\nonumber\\
\PiXV \, \PiYV &=& \exp\Big(i  \sum_\br \big( \chi^x- \chi^y(\PXV[\br])\big) S^z_\br\Big) \PYV \,  \PXV\nonumber\\
&=&  e^{-i \Upsilon}  \PiYV\, \PiXV, \eea 
where we have used
$\big[\PXV\,,\,\PYV\big]=0$. The overall phase is given by the
operator \bea
\Upsilon&=& \sum_\br \omega\ns_\br \, S^z_\br\nonumber\\
\omega\ns_\br&=&  \chi^x+\chi^x(\PYV[\br])-\chi^y-
\chi^y\big(\PXV[\br]\big)\ . \eea

 It can be directly verified from
(\ref{chi}) that \bea \omega\ns_\br = \left\{ \begin{array}{ll}
0&   \br\in \mbox{null lines}\\
2\pi N_\phi & \mbox{otherwise.}
\end{array}\right.
\eea
Since $\exp\big(i2\pi m S^z_\br\big) = (-1)^m$,
\be
e^{-i\Upsilon}= (-1)^{N_\phi  \left( N- N_{\rm
null}\right)}=(-1)^{N_\phi N_{\rm null}}.
\label{Ups-Nnull}
\ee
where $N_{\rm null}$ is the number of sites which sit  precisely on the two null lines.
For  even $N_\phi$, $e^{i\Upsilon}=1$ and $\big[\PiXV\,,\,\PiYV\big]=0$.

For odd $N_\phi$, and odd  $N_{\rm null}$, one obtains  $e^{-i\Upsilon}=-1$.
Let us prove a simple lemma concerning the parity of $N_{\rm null}$.

{\em Lemma}: For  even size  lattices,  if $\bR\ns_\tV$ is tuned to be precisely on a lattice
site, then  $N_{null}$ is  
odd. The proof  is illustrated in   Fig.~\ref{fig:Nnull}.

{\em Proof:} Since we assume that $N$ is even (to describe precise half filling), there are two cases to consider:\\
(i) For an even by even lattice, $\bL=(2m,2n)$, if we choose $\bR\ns_0$ on a lattice site, it is easy top see that  $\bR\ns_\tV$ must also sit  on a lattice site.
The number of sites which contribute to $N_{null}$ are the sum of lattice sites in the $x$ and $y$ directions
minus the null point itself which is counted  twice:
\be
N^{\rm ee}_{\rm null} =2m+2n -1 .
\label{ee}
\ee
Hence $e^{i\Upsilon}=-1$.

(ii) For the odd by even lattice, e.g.  $\bL=(2m+1,2n)$, we choose $\bR\ns_0$, to be in the middle of a bond in the $x$ direction.
The null line includes only the sites on the $x$-null line which is odd:
\be
N^{\rm eo}_{\rm null} = 2m+1
\label{eo}
\ee
Thus, here too   $e^{-i\Upsilon}=-1$. Note  that in both
(\ref{ee},\ref{eo}),  the vorticity center is situated  on lattice
sites  $\bR\ns_\tV=(m,n)$. QED.

Thus we conclude that  for an 
odd number of fluxes   $N_\phi=2n+1$, 
 if $\bR\ns_\tV(\bTheta_i)$ is located precisely {\em on
any lattice site\/}, then
 $\PiXV$ and $\PiYV$  {\em anticommute\/}.

Under these conditions, all states of $\cH[\bTheta_i]$ must be at least two-fold degenerate. This follows  the standard proof:
Since
\be
\left[\cH, \PiXV\right]=0,
\ee
and $\PiXV$ has eigenvalues $\pm1$, then each common eigenstate of $\cH$ and $\PiXV$, can be labelled by $| E_n,\pi^x=\pm1 \rangle$.
Now, 
\be
\PiXV \, \PiYV \, |E_n,1\rangle = - \PiYV \, \PiXV \, |E_n,1\rangle  \propto  |E_n,-1\rangle,
\ee
that is to say each  eigenenergy $E_n$ is associated with a degenerate pair of eigenstates
with opposite  quantum numbers of
$\PiXV$.

$\PiXV$ and $\PiYV$ are point group symmetries about the vorticity center.  We can also construct a third symmetry operator $\Pi^z$ as
\be
\PiZV = -i \, \PiXV \, \PiYV .
\ee
The three operators
$\bPi^\alpha_\tV=(\PiXV \,,\, \PiYV \,,\, \PiZV)$  are unitary and
Hermitian,
 \be
\Pi^\alpha_\tV=(\Pi^\alpha_\tV)^{\dag}=(\Pi^\alpha_\tV)^{-1}\quad\Rightarrow\quad
(\Pi^\alpha_\tV)^2=1.
\ee
Therefore their eigenvalues are $\pm 1$.  The $\Pi$ operators  behave as Pauli matrices and can be used to construct
 an SU(2)  algebra  of spin half,
\be
\tau^\alpha=\half\Pi^\alpha_\tV \quad,\quad \alpha=x,y,z \ .
\ee

We note that for multiple
number of magnetic fluxes $N_\phi>1$,  degeneracies appear for {\em any odd number of
vortices}. This is consistent with the Kramers doublets associated with
an odd number of interacting  spin half particles.

\subsection{V-spin and meron density}
The semiclassical analysis of HCB vortices at half filling, shown in Eq.~(\ref{HF-profile})
finds that the vortex has a CDW in its core.This is
signalled by the  local   order parameter $\langle n^z(\br)\rangle \ne 0$, as illustrated in Fig. \ref{fig:vspin}.

We  define the modified 'meron
density' operator as
\bea
\tau^z(\br) & \simeq  &\nhat \cdot D_x\nhat
\times D_y \nhat,\nonumber\\
D_\alpha&\equiv& \partial_\alpha -iq\bA_\alpha ,
\eea
where $\nhat$ were defined in (\ref{spins}) and
$D_\alpha$ are gauge invariant
derivatives in the $xy$ plane.
In the absence of a gauge field $\bA=0$,
a single {\em
meron} (half a skyrmion)  of a continuos classical field has topological charge
\be
Q= {1\over 4\pi}
\int \!d^2 r~\tau^z (\br,\bA=0)= \pm\half \ .
\ee

 A HCB  representation of $\tau^z_\br$  is constructed  spin-half operators.
In the presence of one flux quantum, we define the modified topological charge operator as
\be
\tilde{Q}= {1\over 4\pi} \sum_\br \tau^z_\br(\bA) .
\ee

 $  \tilde{Q}  $ is not  expected to be quantized at $\pm\half$.
 However, we have found  that in  the  low lying eigenstates of (\ref{xxz}),
its sign correlates  with the eigenvalues  of $\Pi^z$:
\be
\mbox{sign} \Big(\langle E_n, \pi^z | Q |E_n,\pi^z \rangle\Big) = \pi^z .
\ee
We conclude that $\pi^z$
of a single vortex measures the sign of the CDW in its core, with respect to sublattice A.

\section{The Hamiltonian of Quantum Vortices}
\label{sec:VH}
\subsection{Vortex confining potential}
\label{sec: semiclassical potential}
The current density operator is given in the pseudo-spin representation  by
\be
\bj_\bfeta(\br)=  -2iJ q
\left(  e^{iqA_{\br,\br+\bfeta} } S_{\br}^{+}S^{-}_{\br+\bfeta} -{\rm H.c.} \right) .
\label{Jeta}
\ee

By choosing $\bA$ to describe one  flux quantum of uniform magnetic field through the whole lattice one vortex
is introduced into
the low energy eigenstates.  Indeed, we verified that
the exact ground state exhibits a vortex pattern of the current density  $\langle \bj_\bfeta\rangle$, defined in (\ref{Jeta}).
Also, the center of vorticity is agrees with the value of  $\bRV(\bTheta)$ as defined in Eqs. (\ref{eq:vcenter},\ref{Rv}).


\begin{figure}[!t]
\vspace{-0.3cm}
\begin{center}
\includegraphics[width=8cm,angle=0]{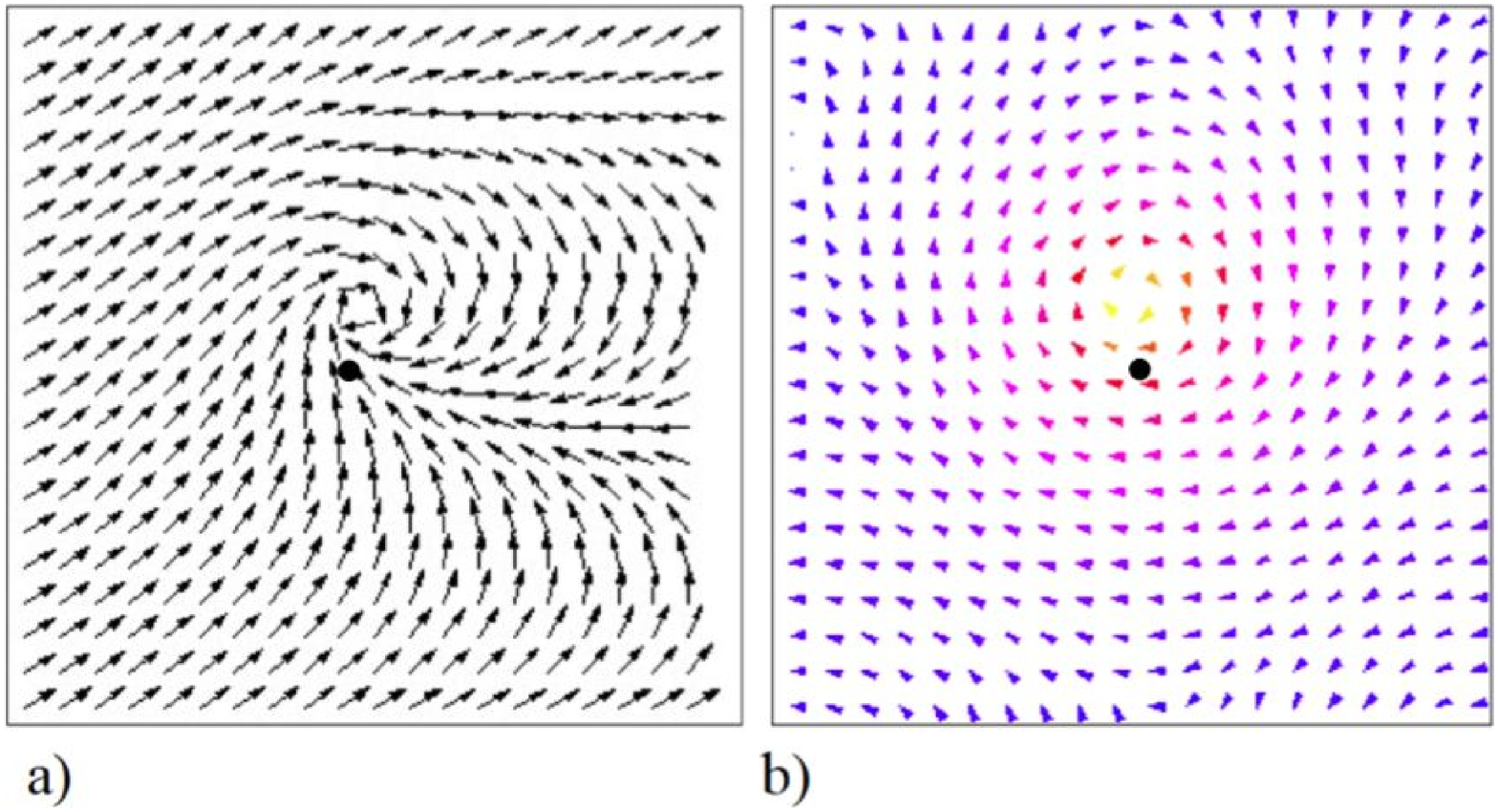}
\vspace{-0.1cm}
   \label{fig:vortex config} \caption{ A variational vortex configuration with its center
located at $\bR=(0.5,0.6)\bL$. The black dots mark the center of vorticity $\bR_{\rm V}$, where
the variational energy is minimized.  a) Phase field marked by directions of the arrows.  b) Current density distribution
where the colors correspond  to the local
current magnitude. Note the excess currents flowing around the
torus in the negative $x$ direction: a consequence of the vortex
being displaced from variational minimum. }
\end{center}
\end{figure}


We determine the effective confining  potential on the vortex variationally. We choose the square geometry $L^2=N$,
and define a  vortex coherent state centered at $\bR$ by a spin coherent state $|\hbOmega\ns_\tV[\bR]\rangle$.
All the unit vectors $\hbOmega\ns_\tV$ lie in the $xy$ plane
with the azimuthal angles $\phi_\br$ given by a Jacobi theta function~\cite{Whittaker}:
 \bea
\phi_\br(\bR)&=& -{\rm Im}\log\vartheta\Big(i(z-Z)-\half-\frac{i}{2}\Big)\nonumber\\
&&\qquad+\pi\Big(y-\big(\half-Y\big)x\Big)\\
\vartheta(z) &=& \sum^{\infty}_{n=-\infty}e^{-\pi n^2}e^{2i \pi n
z}. \label{eq: vortex coherent states}
\eea
Here we use scaled complex coordinates $z=(x+iy)/L$ and $Z=(X+iY)/L$ for the position and vortex center $\br$ and $\bR$,
respectively.

While $\phi_\br$ is discontinuos  on the torus at $x,y=L$,
the  gauge invariant current density
\be
\langle \hbOmega\ns_\tV|  j_\bfeta|\hbOmega\ns_\tV\rangle = q J  \sin\left(\phi_\br-\phi_{\br+\bfeta}-
\bA_{\br,\br+\bfeta} \right) ,
\ee
is  continuous. An example for the phase and current distributions
is given in Fig.~\ref{fig:vortex config}.

\begin{figure}[!t]
\vspace{-0.3cm}
\begin{center}
\includegraphics[width=7cm,angle=0]{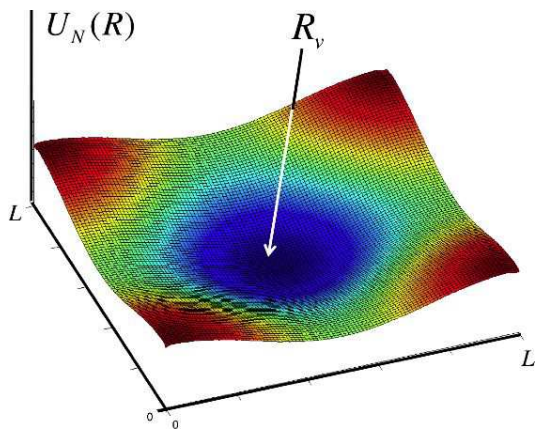}
\vspace{-0.1cm} \caption[The energy as a function of vortex
position]{The energy as a function of vortex position, for a
single vortex on the torus of dimensions $L\times L$.  The
vorticity center is located at $\bRV=\half\bL$.  }
\label{fig:cmpotential}
\end{center}
\end{figure}

The effective confining potential on the vortex is given by the
classical energy defined in (\ref{CSPI}). By semiclassical
estimates \cite{Dan-calc}, its curvature at $\bRV$ scales as
$1/N$. For  lattices of size  $L \geq 4$  we fit the variational
potential by a two dimensional quadratic function, which scales as
$1/N$ as,
\be
U_N = \half K |\bR-\bRV|^2/N, \qquad K=39.2 J.
\label{Un}
 \ee
$U$ is minimized at the vorticity center  $\bRV$, which was
 defined in (\ref{Rv}).  Fig.\ref{fig:cmpotential} depicts the confining potential as a function of vortex center
 for the choice of $\bTheta=(0,0)$.

We can now combine the single vortex hopping  terms of
Eq.~(\ref{Hv}) with the confining potential to obtain the Harper hamiltonian on the finite torus
\bea
H_\tV^N   &=&     - \half t\ns_\tV  \sum_{\bR,\bfeta} \left( e^{i
\cA\ns_{\bR,\bR+\bfeta} } \,  b\yd_{\bR} b\nd_{\bR+\bfeta} + \mbox{H.c.}\right)\nonumber\\
&&\qquad\qquad + U_N(\bR) \,b\yd_\bR b\nd_\bR .
\label{hb}
\eea


\begin{figure}[!t]
\begin{center}
\includegraphics[width=10cm,angle=0]{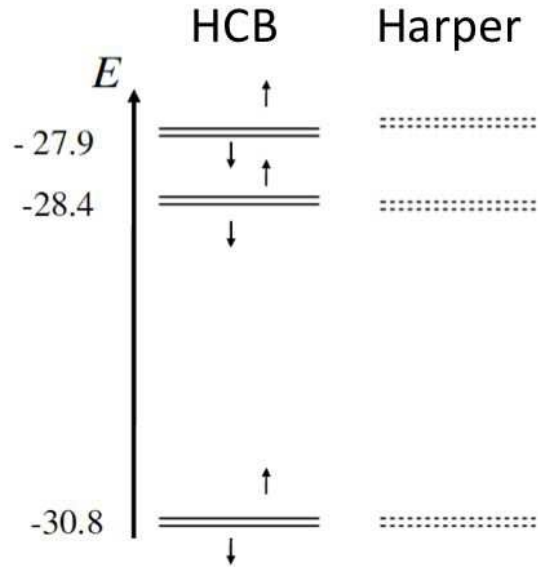}
\caption{Fitting\ the single vortex energies of the effective Harper Hamiltonian (\ref{hb}), to the many-body spectrum of HCB (\ref{xxz}).
The HCB Hamiltonian is defined on a $4\times4$ lattice at half filling, which is embedded on a torus
which is penetrated by one flux quantum. The parameters
$J=1$,  and $V=0$ are chosen for this figure.  Up and down arrows denote the  v-spin magnetic quantum
number $\tau^z$.  The confining potential of the vortex in the Harper Hamiltonian is calculated by
a variational calculation, Eq.~(\ref{Un}). The  effective vortex hopping rate is fit to be $t_\tV = 4J$.
The lowest three doublets of the two models
agree  within $2\%$ of the first energy
gap.}
\label{fig:spectrum}
\end{center}
\end{figure}

\subsection{Vortex hopping Amplitude}
\label{sec: vortex hopping}
For a quantitative quantum theory of vortices we need to evaluate
the effective hopping $t\ns_\tV$. Since vortex tunneling between
lattice sites depends on short range many-body correlations, we
extract $t\ns_\tV$ from exact numerical diagonalizations of $\cH$ on $16-20$ sites  clusters, in the presence of a single flux quantum.

By tuning $t_\tV$, we fit the lowest three
eigenenergies $E_n$  of $\cH$ to those of the effective Harper
Hamiltonian (\ref{hb}). The fit is shown in Fig.\ref{fig:spectrum}.

Our primary concern is that the low eigenstates will not be exclusively described by $H_\tV^N$,
since there are also low energy superfluid phonons (phase fluctuations)
\cite{FL89,ARO97,AA08,MacDonald}.
However, we can estimate the phonons lowest excitation
to be gapped by the finite lattice with the
energy scale $ 2\pi J /L$, which is larger than the energies we have fitted to (\ref{hb}).

Our results for $t\ns_\tV(n\ns_{\rm b}, V/J)$, for $N=20$ can be described by the fitting
 formulas,
\bea
t\ns_\tV  (n,0)  &=&  4J   -50.4 J \left( n-\half\right )^2 + 5056 J \left( n-\half\right )^4,\nonumber\\
t\ns_\tV  (\half,V)  &=&   4J  + 6 V + 10.8 \,{V^2\over J}.
\label{tv}
\eea
The system parameters were varied throughout the range $|n -\half
| \le 0.2$, and  $V/J< 0.5$. We find that at half filling,  the
vortex hopping rate $t\ns_\tV$ varies very little between the $N=16$
and $N=20$  lattices. This  indicates that the bare vortex kinetic
energy is determined by short range correlations, and thus does
not require a large lattice to be computed with acceptable
accuracy.

\begin{figure}[!t]
\begin{center}
\includegraphics[width=8cm,angle=0]{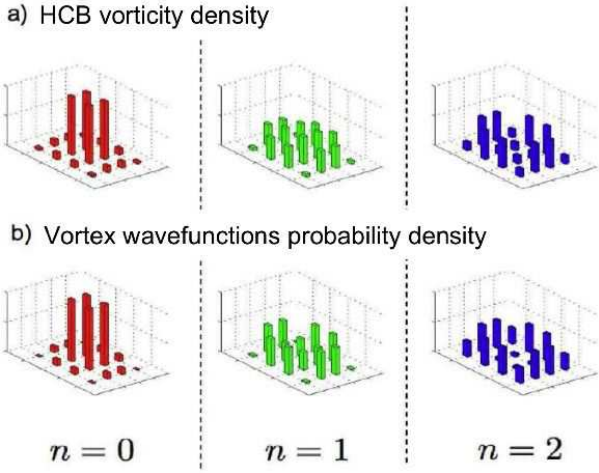}
\caption {(a) The HCB vorticy density $\rho_{\rm v}^{(n)}(\bR)$ as defined in Eq.~\ref{VD}), for the states  $|n,\uparrow\rangle$
whose spectrum is depicted in   Fig.~\ref{fig:spectrum}.
 A uniform background vorticity has been
subtracted. (b) Probability density of the lowest
single particle states of the Harper model given by  Eq.~(\ref{hb}).
The qualitative similarity between (a) and (b) supports the fit of the vortex hopping
rate, and the validity of the Harper Hamiltonian for single vortex dynamics. \label{fig:wavefunctions}}
\end{center}
\end{figure}


To further test the association of the many body lowest eigenstates
with vortex center fluctuations, we   measure the  vorticity density  defined on the dual lattice,
\be
\rho_{\rm v}^{(n)}(\bR)\equiv \sum_{\br,\bfeta}^{{\rm plaq},\bR} \langle \Psi_n|  \bj_{\br,\bfeta}|\Psi_n \rangle \cdot \bfeta ,
\label{VD}
\ee
 for each of the low lying states $|\Psi_n\rangle$. We compare $\rho_v$ to  the single particle
 probability density of the corresponding
wavefunctions of $H^N_\tV$. As shown in
Fig.~\ref{fig:wavefunctions}, for the fitted value of $t\ns_\tV$,
the corresponding distributions increase in width in a
qualitatively similar fashion. This demonstrates that the low
lying eigenstates of $\cH$ correspond to quantum fluctuations of
the vortex position.

\section{Quantum Melting of the vortex lattice}
\label{sec: quantum melting}
At half filling, for $N_\phi>1$  semiclassical evaluation of vortex interactions at distances larger
than the core radius is
\be
U^{\rm int}_{ij} = - \pi J \log(|\bR_i-\bR_j|) .
\ee
 Integrating out the phonon
fluctuations,  produces an instantaneous logarithmic (2D Coloumb)
interaction between vortices, plus retarded (frequency-dependent)
interactions \cite{ARO97,AA08}.  Since we are interested in the
short wavelength fluctuations which are responsible for quantum
melting of the vortex lattice, we ignore these retardation
effects.

Thus in the large lattice limit, at a finite magnetic field $B$,
the multi-vortex quantum hamiltonian is given by the  Harper boson plasma (HBP)
\bea
H\ns_{\rm HBP}  &=&    - \half t\ns_\tV  \sum_{\langle ij\rangle}
\left( e^{i\cA\ns_{ij} } \,
b\yd_i b\nd_j + \mbox{H.c.}\right)\\
&&- \pi  J \sum_{i,j}   n\ns_i\,n\ns_j\, \log(|\bR_i-\bR_j|)
\nonumber\\
&& + \pi^2  J {B\over \Phi_0} \sum_i n\ns_i\, |\bR_i|^2\nonumber
 \label{hbp}
\eea

At half filling the continuum limit of (\ref{hbp}) can be taken as follows. The single vortex dispersion exhibits  a two-fold
degeneracy of the ground states of the Harper Hamiltonian at $\pi$ flux per plaquette.
This implies the degeneracy  $E_{\bk}= E_{\bk+(0,\pi)}$, at low $|\bk|<<\pi$.
Since we wish to expand the hamiltonian at long wavelength, we retain the degeneracy
by the v-spin label  $s=\uparrow,\downarrow$. The vortex effective mass is defined as
\be
{M\ns_\tV}= \left( {\partial^2 E_\bk \over
\partial \bk^2} \right)^{-1}  ={1/t\ns_\tV } .
\label{Mv}
\ee
This leads to the continuum spin-half   Coulomb Bosons (CB) Hamiltonian for the vortices as half filling:

\bea
\cH_{CB}  &=& \sum_{i, s=\uparrow\downarrow}    { \bP_i^2 \over  2{M\ns_\tV} } +  {\pi J}\sum_{i\ne j}
\log(|\bR_i-\bR_j|)  \nonumber\\
&& \qquad-{B\over \Phi_0} {\pi^2 J}  \sum_i |\bR_i|^2 .
\label{CB}
\eea

At low vortex densities, interactions clearly must dominate over
the kinetic energy, and vortices  form a vortex lattice. This a
superfluid  phase with  v-spin correlations.

At a finite temperature  $T_m$,  if we ignore the quantum effects of the kinetic term,
the classical melting temperature $T_m$ is
independent of  vortex density (magnetic field)~\cite{VLmelt1,VLmelt2}.
(Any change in $B$ can be absorbed by scaling  $\bR_i$
 appropriately, leaving the classical energy invariant).

However, quantum fluctuations of the Coloumb plasma
Eq.~(\ref{CB}), increase with the vortex density $B/\Phi_0$,
until quantum melting is reached  at a critical vortex density $
B_{\rm cr}$. This melting is analogous to that of spinless CB,
studied by Magro and Ceperley (MC)~\cite{margo} by diffusion
Monte-Carlo.

MC used  the
dimensionless parameter to describe the CB density $n\ns_{\rm v}$,
\be
r_s^{-2} = \pi n\ns_{\rm v} a_0^2 ,
\ee
where $a_0$ is the Bohr radius.
We set the Bohr radius to be
\be
a_0=\bigg({\hbar^2\over \pi J {M\ns_\tV}}\bigg)^{1/2}
\ee
to match between the model of MC and our  $\cH\ns_{\rm CB}$ of Eq.~(\ref{CB}).

MC
found that below
$r_s
\approx 12$ the boson lattice undergoes quantum melting, i.e.  they found the critical quantum melting density of
\be
n_{\rm v}^{\rm cr} = {1\over 144\pi  a_0^2} .
\ee
Above this density, the CB looses translationally symmetry breaking and becomes a quantum liquid,  which will be  discussed in Section \ref{sec:EXP}. Using our values of $M_{ v}$ from Eqs. (\ref{tv},\ref{Mv}), this
translates into a critical  vortex number per lattice site of
 of
 \be
n_{\rm v}^{ cr} \le \left(6.5-7.9 {V\over
J}\right)\times10^{-3} ~\mbox{vortices per site}.
\label{ncr}
 \ee
This is  a suprisingly low vortex density, above which a vortex quantum liquid (QVL) is created.

\section{Hall Conductivity}
\label{sec:HALL}
We have shown earlier   by Eqs.~(\ref{GI-sxy} , \ref{phs}) that the Hall conductivity
in the low and high density obey the effective Galilean invariant limits,
\be
\sxy =\left\{
\begin{array}{lr}
\frac{n q  }{B}, &n \ll \half\\
- {(1-n)q\over B}~~&1-n \ll \half
\end{array} \right.
\label{sxy-n}
\ee
In terms of vortex motion, this relation implies that below and
above half filling vortices drift in opposite directions relative
to the particle current. In the following we shall study the
transition between these two regimes. Since the continuum
approximation is expected to fail near  half filling, we resort to
a numerical computation of $\sxy$.
 \begin{figure}
\begin{center}
\includegraphics[width=8cm,angle=0]{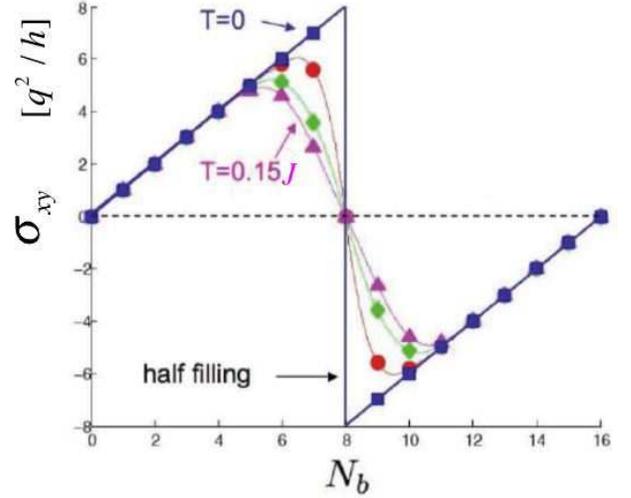}
 \caption{Hall conductance as a function of  boson number
$N_b$ for hard core bosons on a $4\times 4$ lattice on the torus
 with one penetrating flux quantum.
Temperatures vary in intervals of $\Delta T= 0.05 J$. The jump of
the zero temperature conductance at half filling, is smoothened at
finite temperatures.\label{fig:hall} }
\end{center}
\end{figure}

The zero temperature Hall conductance of a finite lattice embedded on a torus is defined by the Chern number~\cite{yosi}:
\be
\sxy(N)= {q^2\over h \pi  }  \int\limits_0^{2\pi}\!\!d\Theta\ns_x\!\!\int\limits_0^{2\pi}\!\!d\Theta\ns_y\>
  \mbox{Im}\left\langle{\partial \Psi_0
\over\partial  \Theta\ns_x}\bigg| {\partial \Psi_0 \over\partial
\Theta\ns_y}\right\rangle ,
\label{sxy0}
\ee
where $|\Psi_0(\bTheta) \rangle$  is the exact
ground state of (\ref{xxz}), in the presence of AB fluxes $\bTheta$. $q^2/h$ is the quantum of conductance.
In the absence of degeneracies and level crossings, the Chern number
$\sigma_{xy}(N)h/q^2 $ is an integer. We  compute Eq.~(\ref{sxy0}) for a sequence of finite lattices.
In Figure \ref{fig:hall} we plot  $\sxy(N)$ for a square lattice of size  $4\times 4$ with aa single flux quantum $N_\phi=1$,
as  function of boson numbers $N_b=[ 0,1,\ldots ,16]$.
We find that the Hall conductance  follows two straight lines as given by Eq.~(\ref{sxy-n}), with an abrupt  jump  to zero at half filling.
The same behavior was found for all smaller  lattices, and reflects a sharp change in vortex dynamics
around half filling.

We extend Eq.~(\ref{sxy0})  to finite temperatures by thermally averaging over
all eigenstates $|\Psi_n\rangle$,
\bea
\sxy(T) &=&  {q^2\over h \pi  } \sum_{n=0}^{\infty} \int\limits_0^{2\pi}\!\!d\Theta\ns_x\!\!\int\limits_0^{2\pi}\!\!d\Theta\ns_y\>{e^{- E_n/T }\over Z} \nonumber\\
&&\qquad\times  \mbox{Im}\left\langle{\partial \Psi_n
\over\partial  \Theta\ns_x}\bigg| {\partial \Psi_n \over\partial
\Theta\ns_y}\right\rangle .
\label{sxy-T}
\eea
$E_n(\bTheta)$  and $|\Psi_n(\bTheta) \rangle$  are the exact
spectrum and eigenstates of (\ref{xxz}).
The results obtained with Eq.~(\ref{sxy-T}) are matched at high
temperatures with the the conductivity calculated using the   Lehmann representation
of the Kubo formula~\cite{Mahan},
\bea
\sxy(T)&=&\lim_{\omega\to i 0^+}   {i \over N Z\omega } \sum_{m,n} {e^{-\beta E_m}-e^{-\beta E_n} \over
E_n-E_m + \omega }\nonumber\\
&&\qquad\times \langle\Psi_n
|\sum_\br j_x(\br)|\Psi_m\rangle\langle\Psi_m|\sum_{\br'} j_y(\br')|\Psi_n\rangle,\nonumber\\
\label{eq: Lehmann}
\eea
where the  current operator $\bj_\bfeta$  is defined  by (\ref{Jeta}).

\begin{figure}[!t]
\begin{center}
\includegraphics[width=8cm,angle=0]{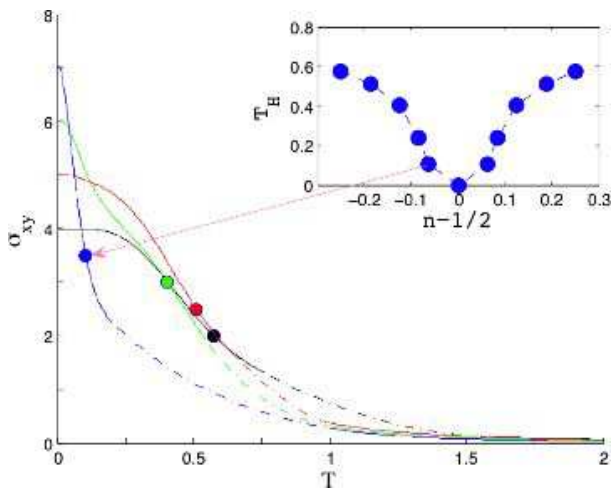}
\caption[Hall conductance of hard core bosons, as a function of temperature]{$\sxy$ of hard core bosons,
as a function of temperature (in units of $t_\tV$), for 4 to 7 bosons on a $4 \times 4$ lattice.
For low temperatures, we calculate $\sxy$ using Eq.~(\ref{sxy-T}),
while for $T>1$ we use Eq.~(\ref{eq: Lehmann}). The dashed line is
an interpolation between the two calculations. The points where
$\sxy$ drops to half its value at $T=0$ are indicated. The inset
shows the temperature scale $T_H$ as a function of density
difference from half filling.}
\label{fig:hall temp}
\end{center}
\end{figure}

The Kubo expression for $\sxy$  is evaluated at high
enough temperatures where the $\omega\to 0$ limit is well behaved.
In Figure~\ref{fig:hall temp} we plot $\sxy$ as a function of
temperature at different HCB densities, by interpolating between  Eq.~(\ref{sxy-T}) at low temperatures
and Eq.~(\ref{eq: Lehmann}) at high temperatures.

We see that in general, the magnitude of $\sxy(T,n\ns_{\rm b})$
decreases with temperature, and the discontinuity  as a function of filling at zero temperature
smoothens at finite temperatures.
As the temperature is lowered, the
reversal of the Hall conductance takes place in a narrower
region around half filling. A characteristic Hall temperature
$T_{H} (n\ns_{\rm b})$ can be defined by,
\be
\sxy(T_{ H}) =\half \,
\sxy(0).
\ee
 In the inset of Fig.~\ref{fig:hall}, we show that $T_{\rm H}$
increases with $|n\ns_{\rm b}-\half|$, although we cannot
estimate the critical exponent from the small cluster  calculation.


\section{Discussion and Experiments}
\label{sec:EXP}
\begin{figure}[!t]
\includegraphics[width=8cm,angle=0]{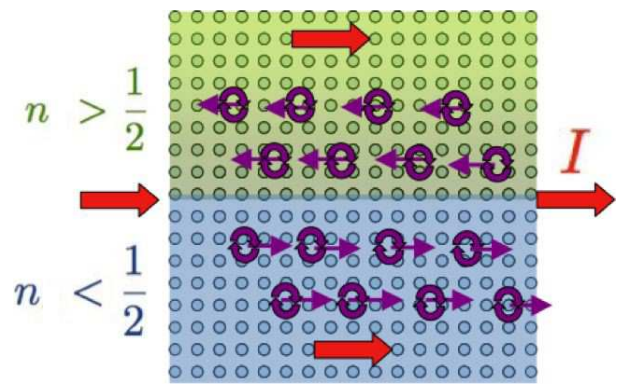}
\caption{Different vortex drift directions (purple arrows)  in the
presence of a bias current (red arrows), for regions of  boson density $n$
which is lower
(blue online) and higher (green online) than half filling.}
\label{fig:Hall-Magnus}
\end{figure}
In this paper,  we have determined the vortex effective hopping
rate $ t\ns_\tV$, in Eq.~(\ref{tv}), and the Hall conductivity
$\sigma_{xy}$, in Fig.~\ref{fig:hall} from the eigenstates of 16
site clusters on the torus. We emphasize that these quantities
serve only as short wavelength ''bare parameters'', to be used to in the
single vortex Harper Hamiltonian Eq.
(\ref{hbp}) and the multi-vortex Bose Coulomb liquid model (\ref{Mv}). The
charge transport coefficients depend on the thermodymamic phases of the
the latter model.

{\em Vortex Solid\/} -- The vortex solid phase, expected for vortex densities lower than $n_{ v}^{ cr} $ of Eq.~(\ref{ncr}),
has  superlfuid (i.e. superconducting) order. Vortices are pinned to their lattice positions and therefore
$\sigma_{xx}=\infty$ and $\sigma_{xy}=0$. The role of $t\ns_\tV$ is to produce quantum zero point motion and effective
v-spin super-exchange interactions, which are ferromagnetic (since vortices have Bose statistics).  Below the v-spin ordering temperature $T_{\rm vspin}$, charge density
waves might be expected with significant magnitude in the vortex cores.  This phase is a weak  supersolid.
Incidentally, density modulations
have been observed near vortex cores of High Tc cuprate
superconductors \cite{v-cores} and analyzed within bosonic models
\cite{SC, Subir}.

At low vortex density, the superexchange interactions decay rapidly, which reduces $T_{\rm vspin}$.
For $T>T_{\rm vspin}$  the v-spins will  contribute an extra  entropy density given by
\be
S_{\rm vspin} = {B\over \Phi_0} \log 2.
\ee

{\em Quantum  Vortex Liquid\/} -- At vortex densities which exceed
$6.5\times 10^{-3}$ vortices per lattice site, we expect the
vortex lattice to melt and give way to the Boson Coulomb Liquid
studied by MC~\cite{margo}.  Superfluidity of the CB translates
into a Mott insulating behavior of the original
bosons~\cite{FL89}. However,  MC have found  that the liquid phase
Eq.(\ref{CB}) is incompresible and hence   exhibits  vanishing
condensate fraction~\cite{NB}. Furthermore, retardation effects
act to suppress dual superfluidity \cite{fiegelman}. The  value of
the transport coefficients of the  QVL phase is therefore left as
an important open question.   Away from half filling, our results
for $\sigma_{xy}$ show that the vortices are subject to a strong
magnetic field, which further suppresses their condensation. At
low boson fillings and large vortex density, $n\ns_{\rm
b}/n_{\phi}<1$, there is  evidence for fractional quantum hall
phases~\cite{sorensen, demler}.

 The QVL phase discussed above is distinct  from the   vortex-antivortex condensate (VC) phases
which were predicted at  rational boson filling fractions, $n\ns_{\rm b}=p/q$, in the absence of a magnetic field~\cite{Lannert,Subir,Zlatko1,Zlatko2}.
These are expected at strong longer range interactions $V\approx J$, and corresponds to  Mott-insulating commensurate CDW phases.

{\em Hall coefficient\/} -- The abrupt reversal of Hall coefficient was found for 16 and 20 site lattices. This effect  correlates with the rapid change in the semiclassical vortex core profile at half filling,
since we know by the GP equation (\ref{GP-profile}) that vortices have a diverging density depletion (accumulation) in the low  and (high) filling regime, while they have  a
localized charge density wave in their small core at half filling. The sign of $\sigma_{xy}$ determines the drift direction of a vortex with respect to a bias current.
This  rapid reversal of $\sigma_{xy}$  may   be relevant to the rapid change in Hall resistivity as a function of doping, which was observed in (La$_{1-x}$Sr$_x$)$_2$CuO$_4$~\cite{Hall-Exp1,Hall-Exp2}.

In Fig.~\ref{fig:Hall-Magnus} we propose a set-up to observe the Magnus action reversal for cold bosonic atoms on a rotating optical trap~\cite{Rotating}. If the density of bosons is allowed to vary
slowly  in space across half filling,
we expect a rapid change of the vortex drift directions at the half filling line. The  vortices would drift downstream  with the boson current  for $n<\half$  and upstream for $n>\half$.

{\em Soft core interactions\/} -- We have not considered relaxing the hard core constraints of the bosons, which would be described by the Bose Hubbard model (\ref{BHM}).
For $U/J <\infty$, the charge conjugation operator ceases to be an exact symmetry at half filling, and  Eq.~(\ref{Part-hole}) is not valid. 
Therefore the Hall coefficient will not be precisely zero at half filling, and the  
v-spin degeneracies will be lifted by the finite $U/J$ corrections.  A full determination of the Hall conductivity in the $U/J$ versus $n$ phase diagram, is an interesting open question.
In particular, one would like to find out  how the zeros of the Hall conductivity connect between the HCB limit and  the free bosons limit.

\begin{figure}[!t]
\begin{center}
\includegraphics[width=8cm,angle=0]{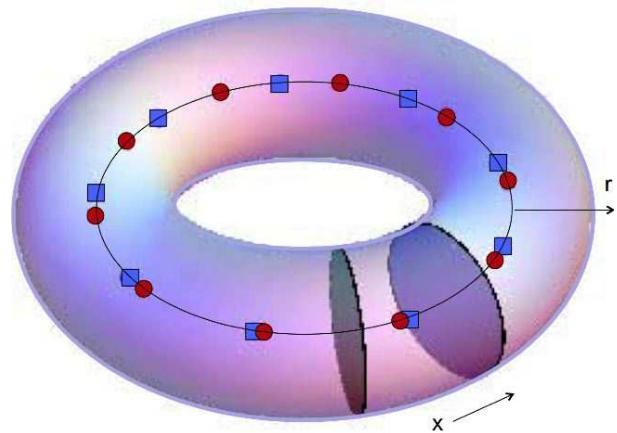}
 \caption{{\em Stern's construction.}  A large number,   $N$,  of positive magnetic monopoles (red circles) and $N-1$ negative monopoles (blue squares)
 uniformizes the  the radial magnetic field, as shown in Fig. \ref{fig: field phi}.  However, the internal flux which is measured
through the disks  which cut through the torus,   oscillates wildly as a function of azimuthal direction $x$, as shown in Fig.\ref{fig: flux}.
}
\label{fig:torus-mon}
\vspace{-0.3cm}
\end{center}
\end{figure}

\subsection*{Acknowledgements}
We thank  Ehud Altman,  Yosi Avron, David Ceperley,  Misha
Feigelman, Steve Kivelson, Gil Refael and  Ady Stern for useful discussions.
Support of the US Israel Binational Science Foundation and Israel
Science Foundation are gratefully acknowledged.  AA and DPA
acknowledge  Aspen Center For Physics where many of the ideas
were conceived. NL acknowledges the financial support of the
Israel Clore foundation.

\appendix
\section{Translational symmetry breaking on the continuous torus}
\label{App:TSB}
In section~\ref{sec:GT} we have shown that
the ground state of the HCB Hamiltonian Eq.~(\ref{xxz}) with
$0<N_\phi<N$ exhibits translational symmetry breaking (TSB) relative to the lattice periodicity,  in both $x$ and $y$ directions.
At first thought this is very surprising: the
flux per plaquette is uniform, so one might expect that physical observables in any non-degenerate state
would not distinguish one lattice point over another.  However, one empirically finds, for example for $N_\phi=1$, that the ground state  current  circulates around a preferred position,  $\bRV$.

\begin{figure}[!t]
\vspace{-0.3cm}
\begin{center}
\includegraphics[width=8cm,angle=0]{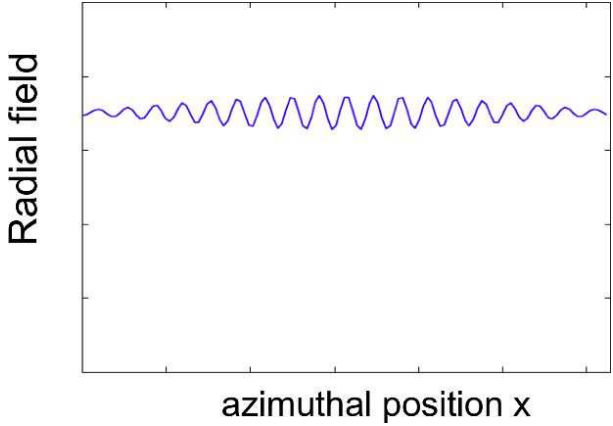}
 \caption{The radial component of the magnetic field in the plane $y=0$,
containing the ring  $N=20$ monopoles and $N=19$ anti-monopoles on the line $y=0$, as shown in Fig.\ref{fig:torus-mon}.
as a function of the azimuthal coordinate $x$. The oscillations magnitude decays as a function of $N$ leading to a uniform
radial  field.}
\label{fig: field phi}
\end{center}
\end{figure}

In this appendix  we  first explain the reason for the TSB. By working out a specific proposal for trying to construct  a purely radial magnetic field, we learn why TSB is
in fact  {\em unavoidable}  on the torus.

In the  particular choice of gauge field in Section \ref{sec:GT},
the gauge invariant Wilson loop functions $W_x, W_y$ defined in
Eq~(\ref{eq: wilson loops}) are linear functions of  $x$ and $y$
modulo $2\pi$. Therefore, by construction  they  break
translational symmetry in  both directions,  and we use them to
define the special null point on the torus, $\bR_0$.

Now we show that by Stoke's theorem, if $\bA$ is continuous on any
interval $[x_1,x_2]$, then $W_y(x)$ must be piecewise linear,
\beq
 W_y(x_2)-W_y(x_1)= qB L_y (x_2-x_1) ,
\label{eq: flux x}
\eeq
where $L_y$ is the circumference of the torus in the $y$
direction. $\bA(\br)$ and $W_y(x)$ however cannot be continuous
everywhere on the torus, since   $W_y(x_2)$ must be periodic for
$x_2\to x_2+L_x$ which is inconsistent with a continuous linear
behavior given by (\ref{eq: flux x}).

The magnetic field of $N_\phi=1$ enters the torus and must end in a
magnetic charge. By Dirac quantization~\cite{Dirac},  'magnetic  charge' density  must be quantized as    $\rho_{m}= g_m\, \delta(\br-\br_m)$ where,
\be
g\ns_{\rm mon}=1/(2q) .
\ee

Therefore the  physical  reason for  the TSB is the point-like
discreteness of the magnetic charge. Consider an embedding of the
torus in three dimensions. Dirac quantization forces one to choose
a special location $\br_m$ inside the torus which breaks the
translational symmetries.

\begin{figure}[!t]
\vspace{-0.3cm}
\begin{center}
\includegraphics[width=8cm,angle=0]{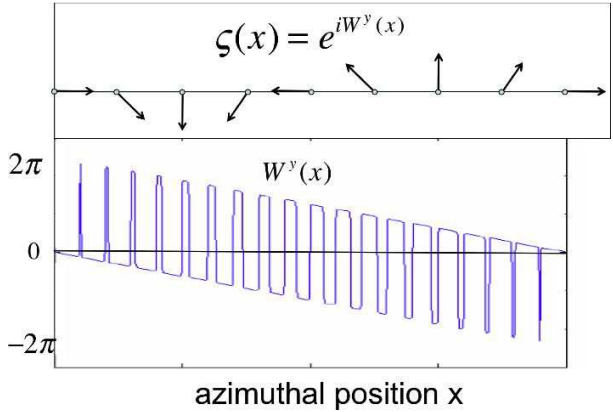}
\caption{Wilson loop function $W_y(x)$, and the phase function $\zeta(x)$,   for the monopole configuration of Fig.~\ref{fig:torus-mon} for $N=20$.
While $W_y$ oscillates wildly at large $N$ and does not converge to a limit, the physically relevant function $\zeta$ is continuous for all $N$ and  breaks translational symmetry. }
\label{fig: flux}
\end{center}
\end{figure}

A possible counter argument is raised: Could  TSB be avoided by
somehow  {\em smearing} the monopole charge inside the torus? This
would presumably restore translational symmetry, at least for the
direction in which the embedding has axial symmetry. It is
somewhat surprising that the answer is negative, as the particular
construction below demonstrates.

Ady Stern~\cite{Stern-comm} has suggested the following
construction. Consider a large number $N$ of positive Dirac
monopoles and $N-1$ negative monopoles placed on the middle circle
inside the torus, as shown in Fig. \ref{fig:torus-mon}. One
arranges  both positive and negative monopoles to be at equal
distances between them such that their mean density is uniform.
Let us calculate the magnetic fields and their effects on bosons
on the torus surface,  and then take $N$ to infinity.

For large $N$, the distribution of monopoles
approaches a uniform density with total
monopole charge of $g_m$. We  compute numerically the magnetic field
created by $N=20$ monopoles situated on rings of radius $R$, as
given by Coloumb's law
\beq
{\bf B}={ \br \over 2qr^3}.
\eeq
The radial magnetic  field penetrating  the surface of the torus in the plane $y=0$,
 becomes  increasingly closer to a constant, with decreasing oscillatory component as shown in Fig.~\ref{fig: field phi}.
 This behavior is precisely analogous to the electric field from a ring with uniform charge density.

Now let us examine the function $W_y(x)$,  evaluated for the same
monopole configuration as Fig.~\ref{fig: flux}. By Gauss's law, it
exhibits discontinuous jumps of size $ 2\pi$  ($-2\pi)$ at each
position where the cross section of the torus at $x$ cross through
a monopole (anti-monopole). Note that $W_y(x)$ exhibits $N$
positive jumps and $N-1$ negative jumps, corresponding to the
number of positive and negative monopoles. Therefore, in order for
$W_y(x)$ to be periodic in $x\to x+L_x$, the continuous part of
$W_y(x)$ needs to compensate for the extra positive jump. Indeed,
as can be seen by Fig.~\ref{fig: flux}, between the $ 2N-1$ jumps
$W(x)$ decreases linearly, as demanded by Eq~(\ref{eq: flux x}).

This increasingly discontinuous function  does not  converge to a
well defined limit function in the large $N$ limit. However, the
physically relevant function  which effects the dynamics of our
bosons of charge $q$ on the surface  is the unimodular phase
function
\be
\zeta(x) = e^{iW_y(x)} = \zeta_0\, e^{iqBL_y x},~~~x\in[0,L_x] ,
\ee
which is perfectly continuous and periodic on the circle. Here we
see that $\zeta(x_0)=1$  uniquely defines a special position $x_0$
which breaks lattice translational symmetry.

We note that  the values of $\zeta(x)$ have physical consequences
on the current distribution.  In the  ground state,   the loops in
the region of   $\zeta\approx 1$ feel a weak  AB flux, and thus a
relatively weak  persistent current is induced in these regions.
Similarly,  the  persistent currents are expected to be maximized
around  loops with $\zeta \approx -1$.

Thus we learn that a static configuration of monopoles
leads  translational symmetry breaking. However,  if one considers the possibility of an extended  quantum {\em wavefunction} of a monopole,  the
magnetic field will be in a quantum superposition. In this case, translational symmetry can be restored
in the entangled  state of the matter field with the  electromagnetic
field.


\begin{thebibliography}{99}
\bibitem{comm:CCS}We note that in the case of half filled fermion bands,  particle-hole conjugation is obtained by sending  $f\ns_i\to f\yd_i$.  However, due to the anticommutation
of   fermion operators, Hamiltonians can realize this symmetry only on bipartite lattices. In contrast,
particle-hole symmetry of half filled HCB models occurs on any graph.
\bibitem{NB} We note that a finite superfluid stiffness has been obtained for the CB phase by
 H. Nordborg and G. Blatter, {\sl Phys. Rev. Lett.} {\bf 79}, 1925 (1997).
\end{thebibliography}
\end{document}